\documentclass[12pt]{article}

\usepackage{epsfig}
\usepackage{cite}
\usepackage{amsmath, amssymb, amsfonts}
\usepackage{color}
\usepackage{latexsym}
\usepackage{graphicx}

\def\be{\begin{equation}}
\def\ee{\end{equation}}
\def\ba{\begin{eqnarray}}
\def\ea{\end{eqnarray}}

\def\bdm{\begin{displaymath}}
\def\edm{\end{displaymath}}
\def\la{~\mbox{\raisebox{-.6ex}{$\stackrel{<}{\sim}$}}~}
\def\ga{~\mbox{\raisebox{-.6ex}{$\stackrel{>}{\sim}$}}~}
\def\bq{\begin{quote}}
\def\eq{\end{quote}}

 at 10truept

\newcommand{\Mpl}{m_{\mathrm{Pl}}}

\newcommand{\bea}{\begin{eqnarray}}
\newcommand{\eea}{\end{eqnarray}}

\newcommand{\bi}{\begin{itemize}}
\newcommand{\ei}{\end{itemize}}

\newcommand{\beq}{\begin{equation}}
\newcommand{\eeq}{\end{equation}}
\newcommand{\beqa}{\begin{eqnarray}}
\newcommand{\eeqa}{\end{eqnarray}}
\newcommand{\mpl}{\Mpl}

\def\la{~\mbox{\raisebox{-.6ex}{$\stackrel{<}{\sim}$}}~}
\def\ga{~\mbox{\raisebox{-.6ex}{$\stackrel{>}{\sim}$}}~}

\def\ltap{\ \raise.3ex\hbox{$<$\kern-.75em\lower1ex\hbox{$\sim$}}\ }
\def\gtap{\ \raise.3ex\hbox{$>$\kern-.75em\lower1ex\hbox{$\sim$}}\ }
\def\gl{\ \raise.5ex\hbox{$>$}\kern-.8em\lower.5ex\hbox{$<$}\ }
\def\roughly#1{\raise.3ex\hbox{$#1$\kern-.75em\lower1ex\hbox{$\sim$}}}

\begin{document}

\thispagestyle{empty}
\begin{flushright}
March 2019 
\end{flushright}
\vspace*{1.75cm}
\begin{center}

{\Large \bf Dark Energy,  $H_0$ and Weak Gravity Conjecture}

\vspace*{1cm} {\large Nemanja Kaloper$^{a, }$\footnote{\tt
kaloper@physics.ucdavis.edu}
}\\
\vspace{.3cm} {\em $^a$Department of Physics, University of
California, Davis, CA 95616, USA}\\

\vspace{1.5cm} ABSTRACT
\end{center}
We point out that the physics at the extreme IR---cosmology---might provide tests of the physics
of the extreme UV---the Weak Gravity Conjecture. 
The current discrepancies in the determination of $H_0$ may hint at a modification of $\Lambda$CDM. An  
extension which may fit better comprises of an early contribution to dark energy which `decays' into 
relativistic matter. On the other hand the discourse on WGC to date suggests that fields which support 
cosmic acceleration may produce relativistic matter after they traverse a $\sim$ Planckian distance 
in field space. We explain how this offers a simple realization
of the requisite cosmic phenomenology. Thus if the resolution of $H_0$ discrepancies is really 
early dark energy that ends with a shower of relativistic matter and the current ideas on WGC are 
indicative, this may be a rare opportunity to
link the two extreme limits of quantum field theory.

\vfill \setcounter{page}{0} \setcounter{footnote}{0}
\newpage

Cosmology is the discipline of deciphering the results of a single experiment, which is still in progress. 
Just like in a collider experiment,
we can only go after the outcome of a process at times longer than the interval for the process to complete. Even determining the precise details of the initial and final states is difficult due to the inherent imprecision of the measurement process predicated by time evolution. A single scalar field, for example, can be both dark energy, dark matter and even radiation, depending on the epoch of cosmic evolution. This makes it hard to be certain just what the cosmic contents is. The current fit is the concordance cosmology aka $\Lambda$CDM model, established 
by combining observations of CMB anisotropies, the LSS distributions and SNeIa redshifts and distances. According to it, the universe contains $\sim 70\%$ of vacuum energy, $\sim 26 \%$ of dark matter and $\sim 4\%$ of the particle physics Standard Model matter in a spatially Euclidean, expanding geometry. 

As the quality and the amount of data improve  the model may change again, as it did before. Thus the current 
discourse on the determination of $H_0$ is very interesting \cite{H0}. If the different methods really yield different values of $H_0$ within the same underlying $\Lambda$CDM cosmology, we may need to change $\Lambda$CDM to a more intricate fit, with modifications of the current ingredients or 
new ingredients altogether. Among the many possible extensions clearly those involving dark energy stand out.
If dark energy is a constant, we can only learn very little more about it. If it is not constant, we may get new direct insights
into the dark sector. Curiously, a recent analysis \cite{kamion} suggests that dark energy evolved during the cosmic history, having quickly changed at some redshift well prior to the epoch of recombination. 

To fit the data, however, this variation of dark energy isn't enough by itself \cite{linder,kamion2,lloyd}. Importantly, the extra early dark energy (EDE) should decay and dilute away at least as fast as radiation, or an even faster diluting component. This helps by making cosmic expansion faster early on, and delaying matter-radiation
equality by adding a small amount of radiation. Hence the value of $H_0$ from cosmological measurements
increases, converging with its locally measured value. For this reason, the analysis of \cite{kamion} involves what seems like a more exotic set of dark energy potentials. With typical dark energy potentials with slow roll, once a field falls out of it, it starts to oscillate around the minimum, with energy density scaling like
massive matter due to a vanishing pressure. The oscillations could dissipate via particle production, but in perturbation theory the rate of decay of the field energy is $\Gamma \sim m^3/{\cal M}^2 \ll m$ 
where $m$ is the mass of the field, and ${\cal M}$ the scale this suppresses the mixing of the field and its decay products. If that field was EDE, it must be very light, with $m \la H_c \sim H_0 \times (5000)^{3/2} \sim 10^{-27} {\rm eV}$. Hence
the decay products must be ultralight too, with masses even smaller than $m$. Even then, the lifetime of the field energy would be too long, $\Gamma^{-1} \gg H_c{}^{-1} \sim m^{-1}$, and the energy density after slow roll would
scale as matter for too long. In turn \cite{kamion} consider potentials which aren't simple broad and shallow quadratics near the minimum. This gives the effective equation of state which yields 
redshift rate faster than massive matter. So while the idea is that EDE is an axion like field, one of many  \cite{nessy,kamion3} which dwell in the axiverse \cite{axiverse}, due to a more complicated potential 
\cite{kamion} it retains its energy density, but the self-interactions
change the pressure and yield faster dilution. 

This is an interesting suggestion. However much of the model building of axion dark energy involves 
potentials which are shallow and broad quadratics to the leading 
order \cite{hill,frieman,yanagida,nilles,barbieri,KSQ,hobbit,SCQ}. It would appear that these 
`run of the mill' axion quintessence models may not explain EDE because of how they behave 
after they fall out of slow roll. On their own, they would seem to yield contributions to dark matter rather than radiation, as described above. On the other hand, they are perfectly fine candidates to account for the present dark energy which dominates the universe now. This odd asymmetry looks awkward. 

However UV completions of such models introduce additional ingredients. Their effective field theory (EFT) includes a large number of massive states, whose masses however
depend on the low energy field vevs. Recent discussions of the Weak Gravity Conjecture (WGC) \cite{bdfg,weak,vafao} and its implications for cosmology \cite{gary,arthur,reece,palti,valenzuela,steinvafa,opsv,ibanez} 
suggest that once a field's
vev changes by $\sim \mpl$, a large number of states may become extremely 
light.  On the other hand, the variation
of a field by $\Delta \varphi \sim \mpl$ is precisely how the end of the slow roll regime occurs. A field which is stuck on a potential slope while the Hubble friction is large starts to `thaw' and roll down when the cosmic expansion slows to $H \la m$. After that, the field oscillates around the minimum, and the oscillations excite the large number of light particles. The decay rates can be greatly enhanced in the presence of a large number of light states, 
and even more so when the masses depend on the vev of the oscillating field, due to parametric resonance \cite{linde,beauty}. For these reasons, we want to propose that the outcome of the analysis of
\cite{kamion} is, in our opinion, a very interesting possibility, that

\vskip.3cm
\noindent {\it If an EDE which dumps most of its energy into radiation at some redshift between $3000$ 
and $5000$ is really the resolution of the problems with $H_0$, this might be an observational signal of
WGC.}
\vskip.3cm

\noindent We are fully aware that this is speculative in the extreme. Nevertheless, given the paucity of chances to probe extreme UV by the tools available to us, we feel we would be remiss to ignore this possibility. Let us now flesh out the details. 

`Traditional' axion quintessence proposals start with a harmonic potential arising due to nonperturbative phenomena in some gauge theory to which the axion couples, 
\be
V \sim \mu^4\Bigl(1-\cos\frac{\varphi}{f} \Bigr) \, ,
\label{potharm}
\ee
where the gauge theory goes strong at a scale $\Lambda$, and where nonperturbative effects yield exponential suppression of $\mu \sim \Lambda e^{-S}$. Here $S$ is the euclidean action of a single gauge instanton in dilute instanton gas 
approximation \cite{hill,frieman,yanagida,nilles}. If this is the case, then to realize slow roll and have $\varphi$ stand in for dark energy presently, when it dominates the cosmic expansion, $f \ga \varphi \ga \mpl$, 
both theoretically \cite{frieman} and from the data \cite{liddle}. In turn this means that one needs to arrange for the couplings and scales in the gauge theory such that $\mu^2 \sim \Lambda^2 e^{-2S} \sim H_0 \mpl$, 
and so $\mu^2/f \la H_0$. While this takes some effort, at least the result is stable in perturbation theory: UV effects
cannot contaminate significantly the low energy theory at near the axion mass. 
The problem with this suggestion is that 
it seems $f\ga \mpl$ is at odds with UV complete frameworks \cite{bdfg,weak}. This might even be impossible 
if one insists on a single low energy EFT describing field domains of
super-Planckian sizes \cite{bdfg,weak,steinvafa,vafao,ibanez,reece,palti,valenzuela,opsv}. 

In parallel, models of monodromy emerged to address the challenges with large field displacements and UV sensitivity \cite{irrational,knp,eva,KS}. In those constructions, one expects a large number of additional massive states to appear: KK states from compactifications, states localized on brane sources and so on. Since 
the axions are among the moduli of the compactifications, the masses of the extra states will depend on the axion
expectation values, at least indirectly. The considerations of WGC \cite{vafao,reece,palti,valenzuela,opsv} suggest that this dependence may be such that when an axion is displaced by ${\cal O}(1) \, \mpl$, a large fraction of 
these states may become very light, and some even massless. Thus, the argument goes, once an axion traverses a 
Planckian distance, a large number of states become light and cannot be left out of the EFT, requiring that 
EFT be modified, and even triggering a phase transition. To be really true this requires that at least an ${\cal O}(1)$ fraction of the states must become lighter than the axion \cite{reece}. If they were not, thanks to decoupling 
we could still retain a subregime of the original EFT of the axion as a light mode, add a few more light degrees of freedom and modify the cutoff. A case in point are KK compactifications and EFTs of shape moduli, such as eg. Brans-Dicke scalar coming from a compactification of a 5D gravity on a circle. Other examples show up
in brane constructions, which have large numbers of relativistic states in coincident limits.

While we remain agnostic on the universality of these statements\footnote{There is much debate on what 
${\cal O}(1) \, \mpl$ actually means. This can affect in significant ways whether we can use axions for various phenomenological applications, and is an important  question that
should be resolved (see eg \cite{westy} for recent developments) but in our case, it suffices to have 
\underbar{one} axion for which ${\cal O}(1) \la 1$.}, due to the inherent imprecisions of calculating the numerical coefficients in low energy EFTs emerging from string constructions at this time\footnote{This is where factors of $4\pi$ can make a huge difference, see eg \cite{SCQ,KS} for field theory examples.}, here we will assume that so strong a
form of WGC is true for at least some ultralight axions. We will further assume that it is one of those axions that becomes the EDE quintessence. To parameterize its dynamics, we can use an effective action 
which includes the corrections from many massive states, at the cutoff $M$, which includes in principle
infinitely many irrelevant operators. These can be organized as in \cite{SCQ} using Na\"ive Dimensional Analysis
devised for heavy quark EFT \cite{NDA}. Truncating the EFT to only two derivative terms for simplicity, we have
\be
{\cal L} = -\frac12 {\cal Z}_{eff}(\frac{4\pi m\varphi}{M^2}) (\partial_\mu \varphi)^2 - \frac{M^4}{16 \pi^2}{\cal V}_{eff}(\frac{4\pi m\varphi}{M^2}) \label{eq:twod}\ ,
\ee
where ${\cal Z}_{eff}$ and ${\cal V}_{eff}$ are dimensionless functions including wave function and potential renormalizations due to the virtual massive modes, with dimensionless coefficients of ${\cal O}(1)$. 
The theory may also contain direct couplings to the massive modes\footnote{In fundamental theory these arise because the masses depend on stabilized moduli, and these in turn depend on axion energy, and therefore on axions. Thus our 
field $\varphi$ may have started as a fundamental axion, but with all the corrections from integrating the heavy modes included, it is better to view it as a collective order parameter of the theory. We will drop the label axion from here on.}, eg ${\cal L}_{int} \ni \mu(4\pi m \varphi/M^2) \bar \psi \psi$ for fermions and similar terms for bosons, as well as derivative couplings like $\partial_\mu \phi \bar \psi \gamma^\mu \psi/\mpl$, which are automatically consistent with the EFT of the light axion with a naturally flat potential due to shift symmetry. The derivative terms  can help keep control over the shift-breaking corrections, thanks to the mechanism of field seizing \cite{scottsavas}, where renormalization of kinetic terms compensates the corrections to the potential. Both the mass term and the kinetic term are renormalized by similar factors\footnote{$\delta m^2 \sim \sum_\psi m^4_\psi/\mpl^2 \sim m^2 \sum_\psi (m_\psi/M)^2$, $\delta Z_{eff}  \sim \sum_\psi (m_\psi/M)^2$, since we take  $m_\psi \sim \mu \la M$ and  $m \mpl \sim M^2$, see below.} which therefore can be absorbed away by a change of normalization. Integrating out such terms for heavy modes contributes to (\ref{eq:twod}). However if $\mu$ has
zeros at some values of $\varphi$, integrating out these terms is no longer possible near those values of $\varphi$. 
Instead they provide decay channels for dissipating dark energy into light particles. 
The  WGC obstructions to large field variations in EFT, then, can be modeled by assuming that $\mu$'s generically
have many zeroes, separated by $\Delta \varphi \simeq \mpl$. As we will see shortly this means that
$M^2 \sim H_c \mpl$ and $m \sim H_c$ where $H_c$ is the Hubble parameter at the time EDE began to decay away. 

So let us assume that $\varphi$ is the EDE. We can field redefine the field using
\be
\chi = \int d\varphi \Bigl( {\cal Z}_{eff}(\frac{4\pi m\varphi}{M^2}) \Bigr)^{1/2} \, , 
\label{chi}
\ee
to bring the kinetic term to the canonical form. Now, as explained in \cite{kamion}, EDE in the early epoch cannot be
dominant, but it needs to be significant, comprising almost a tenth of the cosmic contents then. This means, the 
vev of $\chi$ must be large, far from the potential minimum, where it sits in slow roll. These conditions differ
from the situation when $\chi$ dominates over other sources however, as we can easily see from reanalyzing 
slow roll equations in this context. The background equations are
\be
3H^2 \mpl^2 = \rho_{\tt matter} + V(\chi) \, , ~~~~~~~~ 3H \dot \chi + \partial \chi V = 0 \, ,
\label{chieqs}
\ee
where we have dropped the $\dot \chi$ terms in the first equation and $\ddot \chi$ in the second, assuming we 
are in slow roll. The potential is $V = M^4 {\cal V}_{eff}/16\pi^2$ after trading $\varphi$ for $\chi$. Now we require
that $w_\chi = - \frac{1 - \dot \chi^2/2V}{1+\dot \chi^2/2V} \simeq -1 + \dot \chi^2/V$ is close to negative unity, say 
with the precision of at least ten percent, and that $V$ is about a tenth of $3H^2 \mpl^2$ \cite{kamion}. Combining this and (\ref{chieqs}) yields
\be
\partial_\chi V \sim \frac{V}{\Delta \chi} \simeq \frac{V}{\mpl} \, .
\label{slowroll}
\ee
The important lesson from this is that although any subdominant field 
lighter than the Hubble parameter will be in slow roll for any value of its vev, if it is to imitate EDE with an equation of state $w_\chi \sim -1$, be a nonegligible
contribution to $H_c^2$ and decay shortly after $1/H_c$ its vev must be close to the Planck scale (but not much larger!). This means, that the field will imitate EDE for an efold or so when $H \sim m$ and not much longer. 
Thus, as we announced above, $H_c \sim m$. Further, at this time $M^4 \sim H_c^2 \,  \mpl^2$, fixing the cutoff $M$.
In sum, within a Hubble time set by the mass of the field, this EDE will start to oscillate around its minimum with the frequency $\sim m$. 

By itself, these oscillations would have energy density of nonrelativistic matter. The field would decay into other matter, but 
perturbatively the rate would be very small. However, by our assumption, which is motivated by WGC, many
of the heavy fields at this point become light\footnote{A realization of such a scenario can be found in models where 
many branes collapse onto each other as $\chi$ goes through zero, effectively triggering a phase transition. In this way $\chi=0$ is an accumulation point for the extra relativistic states in the theory singled out by, eg enhanced chiral symmetry.}. 
Since the mass terms are normalized to $M$ at the cutoff, $\mu \la M$, expanding them near the minimum, which we choose to be
at $\chi=0$ for simplicity, where the masses vanish, we'd get 
\be
\mu(\chi)  \bar \psi \psi \sim 4\pi \frac{m}{M} \, \chi \,  \bar \psi \psi\sim 4\pi \sqrt{ \frac{H_c}{\mpl} }\, \chi \,  \bar \psi \psi \, .
\label{masses}
\ee
So as $\chi$ passes through zero, it will excite these particles and dump energy much more quickly than in perturbation theory, using parametric resonance \cite{linde}. From Eq. (\ref{masses}), the decay rate per channel \cite{linde} is 
$\Gamma_{\bar \psi \psi} \sim 2\pi \frac{H_c^2}{\mpl}$, which is still very slow (although it is much faster than
the perturbative answer). The same holds for bosons. 
However, the total rate of energy loss is
\be
\Gamma_{\tt total}  \sim \sum_{\tt flavor} \Gamma_{\tt flavor} \sim 2\pi \frac{H_c^2}{\mpl}  \, {\cal N} \, , 
\label{totalrate}
\ee
where ${\cal N}$ is the fraction of the modes which became light. Since the total number of modes is constrained
by the cutoff condition ${\cal N}_{\tt total} \la \mpl^2/M^2$, assuming that this inequality is almost saturated, and
that WGC implies that ${\cal N} \sim {\cal O}(1) {\cal N}_{total}$, as we stated earlier, we finally find the total decay rate formula which is of crucial importance,
\be
\Gamma_{\tt total} \sim 2\pi \frac{H_c^2 \, \mpl}{M^2} \sim 2\pi H_c \, .
\label{totrate}
\ee

As the field tries to move away from the minimum, the light modes fatten up. 
So once the light modes are produced their energy density
counteracts the continuing motion of the field $\chi$ because increasing the mass of the light modes
costs energy \cite{beauty}. This pins the field $\chi$ to the minimum\footnote{This is actually an enhanced symmetry point like in \cite{beauty}; the symmetry is at least the chiral symmetry of the fermions repeated ${\cal N}$ times.} $\chi=0$ rapidly, within a Hubble time
$1/H_c$ or so. As a result, the produced particles remain light and redshift as radiation from this point on. Because
of (\ref{totrate}), this means that almost all of the EDE has been converted into radiation in less than a Hubble time,
$\Gamma_{\tt total}{}^{-1} \sim H_c{}^{-1}/2\pi$, after the field fell out of slow roll. Detailed microscopics of such
processes is further studied in \cite{evaco,calimero}. 

Interestingly, a low energy theory with so many light species yields a prediction for lab searches, 
due to the universality 
of gravity. The ${\cal N}$ ultralight species will run in the loop and correct the short-distance $4D$ Newtonian potential at the scale $1/M \sim {\rm eV}^{-1}$, which is about a micrometer. This well known fact \cite{radko,duff,donoghue} is summarized in 
the leading quantum correction in the potential between two test masses $m_1$ and $m_2$, 
\be
V_N \simeq - G_N \frac{m_1 m_2}{r} \Bigl(1 +  \frac{ {\cal N}}{\mpl^2 r^2} \Bigr) \simeq - G_N \frac{m_1 m_2}{r} \Bigl(1 + \frac{{\cal O}(1)}{M^2 r^2 }\Bigr) \, .
\label{newt}
\ee
Since $M \la {\rm eV}$, these corrections could be accessible to the future searches for deviations from 
Newton's law  \cite{adelberger}.

One may also worry that with such a large number of ultralight species there could be problems with collider bounds. For example, the processes
$p \bar p \rightarrow {\rm missing ~energy}$, which are mediated by $4D$ gravity at the LHC, and hence typically very weak, could be greatly 
enhanced by a factor of ${\cal N} \sim (\mpl/M)^2 \gg 1$.
However at the cm energy $\sim \sqrt{s_{CM}} \sim {\rm TeV} \gg M$ there will be many other 
decay channels available to the virtual graviton which mix in the transition amplitude 
and yield a form factor, that can reduce the rate dramatically. 
To really determine the cross section for $p \bar p \rightarrow {\rm missing ~energy}$ 
a UV completion to scales above $M$ is needed, which can highly suppress the interactions. 
A similar behavior was noted in the context of RS2 braneworlds in \cite{ja}.

This dynamics therefore {\it precisely} realizes the EDE scenario for resolving the $H_0$ tension in \cite{kamion}, once the axion mass is chosen to be\footnote{The analysis of \cite{kamion} takes the critical scale to be between
the redshifts of $5000$ and $3000$, but this only allows for a narrow window of scales since $(5/3)^{2} \sim 2.77$.}
\be
m \sim H_c \sim H_0 \times (5000)^{3/2} \sim 10^{-27}\, {\rm eV} \, . 
\label{massnumber}
\ee
Interestingly, this puts the EDE field in the window of ultralight dark matter which is
constrained by current cosmological observations \cite{barbiamen,pedro}. Thanks to the fact that its energy
density was subdominant to start with, and, crucially, that it dissipated away via ultralight particle preheating, 
which is fast as per Eq. (\ref{totrate}), it can easily pass the current bounds. 
As a result, the dynamical scenario which we have 
proposed here appears to be a viable fundamental answer to the trouble with $H_0$ \cite{H0,kamion}, 
that might however be accessible to further tests in the future, both cosmological and laboratory. We consider this a rather fortunate serendipity.

In summary, we have proposed that resolving the discrepancies in determining the 
value of $H_0$ might shed light on the physics of the extreme UV. The problems with $H_0$ might be resolved by an early dark energy component which decays quickly into radiation. This could be a consequence of the Weak Gravity Conjecture, yielding bounds on low energy physics arising from quantum gravity. The EDE could be an ultralight field which can vary over large ranges in field space, as in monodromy. It couples to many fields
which can become relativistic as the field changes by a Planckian value, as implied by WGC. Such a 
large number of relativistic species provides a reservoir that the field can quickly dump its energy into 
by preheating. The
preheating rates and the number of species fold in just the right way to ensure that the energy loss by the
field is so fast that it really can behave as the EDE should \cite{kamion}. If it is really true that $H_0$ is resolved by the EDE decaying into radiation, this could be an extremely rare chance to study quantum gravity by data analysis,
on the sky and in the lab. In our view this is a unique opportunity to link the fundamental theory to the largest scale experiments, comprising of dark energy surveys. 

\section*{Acknowledgements}

N.K. thanks G. D'Amico, A. Lawrence, M. Montero, J. Moritz, R. Rattazzi, G. Shiu and A. Westphal for useful discussions. 
This work is supported in part by DOE Grant DE-SC0009999.

\end{document}